\documentclass[a4paper]{article}
\usepackage{tikz}
\usetikzlibrary{matrix}
\usepackage{amscd,amsmath,amssymb}
\usepackage{bbm}

\newcommand{\p}[1]{(\ref{#1})}

\newcommand{\cB}{{\cal B}}

\newcommand{\bQ}{{\overline Q}{}}

\newcommand{\bxi}{{\bar\xi}}

\newcommand{\cN}{{ {\cal N}   }}

\newcommand{\tPi}{{\widetilde \Pi}}

\newcommand{\mH}{{\mathbb H}}
\newcommand{\mQ}{{\mathbb Q}}
\newcommand{\bmQ}{{\overline{\mathbb Q}}}

\newcommand{\und}{\qquad\textrm{and}\qquad}
\newcommand{\tr}{\,\textrm{Tr}\,}
\newcommand{\diff}{\textrm{d}}
\newcommand{\im}{\textrm{i}}

\newcommand{\be}{\begin{equation}}
\newcommand{\ee}{\end{equation}}
\newcommand{\bea}{\begin{eqnarray}}
\newcommand{\eea}{\end{eqnarray}}

\newcommand{\ba}{\begin{aligned}} \newcommand{\ea}{\end{aligned}}

\begin{document}

\thispagestyle{empty}
\begin{flushright}
%\today
\end{flushright}\vspace{1cm}
\begin{center}
{\LARGE\bf Integrability of supersymmetric Calogero--Moser models}
\end{center}
\vspace{1cm}

\begin{center}
{\large\bf  Sergey Krivonos${}^{a}$, Olaf Lechtenfeld$^b$ and Anton Sutulin${}^{a}$}
\end{center}

\vspace{0.2cm}

\begin{center}
{${}^a$ \it
Bogoliubov  Laboratory of Theoretical Physics, JINR, 141980 Dubna, Russia} \\[4pt]
${}^b$ {\it
Institut f\"ur Theoretische Physik and Riemann Center for Geometry and Physics, \\
Leibniz Universit\"at Hannover, Appelstrasse 2, 30167 Hannover, Germany}

\vspace{0.5cm}

{\tt krivonos@theor.jinr.ru, olaf.lechtenfeld@itp.uni-hannover.de, sutulin@theor.jinr.ru}
\end{center}
\vspace{2cm}

\begin{abstract}\noindent\large
We analyze the integrability of the $\cN$-extended supersymmetric Calogero--Moser model. 
We explicitly construct the Lax pair~$\{L,A\}$ for this system, which properly reproduces all equations of motion. 
After adding a supersymmetric oscillator potential we reduce the latter to solving $\dot{U}\,{=}\,A\,U$ 
for the time evolution operator~$U(t)$. The bosonic variables, however, 
evolve independently of~$U$ on closed trajectories, as is required for superintegrability.
To visualize the structure of the conserved currents we derive the complete set of Liouville charges 
up to the fifth power in the momenta, for the $\cN{=}\,2$ supersymmetric model. The additional, 
non-involutive, conserved charges needed for a maximal superintegrability of this model are also found.
\end{abstract}


\vskip 0.5cm


\newpage

\setcounter{equation}{0}
\setcounter{page}1
\section{Introduction}
The original rational Calogero model of $n$ interacting identical particles on a line~\cite{Calogero} is  given by the classical Hamiltonian
\be\label{CH}
H\ =\ \frac{1}{2} \sum_{i=1}^n p_i^2 + \frac{1}{2}\sum_{i\neq j} \frac{g^2}{\left( x_i{-}x_j\right)^2}\ .
\ee
This model receives much attention in different branches of physics such as high-energy and condensed-matter physics. 
{}From a mathematical point of view, the Calogero--Moser model, as well as its variants with special potential terms,
belongs to an important class of integrable and even superintegrable systems (see, e.g.,~\cite{OP}).

Unsurprisingly, the Calogero--Moser model has often been the subject of ``supersymmetrization'', 
beginning with the $\cN{=}\,2$ supersymmetric model of Freedman and Mende~\cite{FM}.
However, all attempts to construct $\cN{=}\,4$ supersymmetric extensions, despite the announced importance of such models~\cite{GT}, 
were unsuccessful due to a barrier encountered in~\cite{Wyll,BGL}. 
To surmount this barrier new supersymmetric Calogero-like models have been proposed in~\cite{scm8,nCal3,nCal2,nCal1}. 
Finally, a supersymmetric Calogero--Moser system with arbitrary $\cN$-extended supersymmetry has been constructed~\cite{KLS1,KLPS}.

The main feature of the latter models is an increased number of fermionic coordinates, namely $\cN n^2$ rather than the $\cN n$ to be expected.  
It is therefore questionable whether they inherit the (super)integrability of the bosonic Calogero--Moser model.
To settle this issue we must first determine how many conserved currents are required for Liouville or super-integrability
in a supersymmetric model with $n_{\text{bos}}{+}n_{\text{fer}}$ degrees of freedom.
Recall that, in the standard Lax description with a pair $\{L,A\}$ of matrices subject to $\dot{L}=[A,L]$, the Liouville charges appear as
the trace of powers of the $L$~operator. In the supersymmetric extension, the Lax pair still produces all (bosonic and fermionic) equations of motion
and $n_{\text{bos}}$ Liouville currents as before, but it is unclear how additional $n_{\text{fer}}$ conserved charges may arise and whether 
they should be commuting or anticommuting in nature. 
Of course, the problem extends to any additional (non-involutive) conserved charges, as required for superintegrability.

In this paper we analyze the integrability of the $\cN$-extended supersymmetric Calogero--Moser model.
For convenience we add a confining supersymmetric oscillator potential.
We explicitly construct the Lax pair for this system and demonstrate that it yields all (bosonic as well as fermionic) equations of motion. 
Employing the Olshanetsky--Perelomov approach~\cite{OP} together with an observation by Wojciechowski~\cite{sCal}
we solve the bosonic equations of motion. Their periodic trajectories prove the maximal superintegrability of this sector.
 
As a definition of (super)integrability for a supersymmetric system with $n_{\text{bos}}{+}n_{\text{fer}}$ degrees of freedom 
we adopt the formulation of Desrosiers, Lapointe and Mathieu~\cite{DLM1,DLM2}:\footnote{
For a different definition, see e.g.~\cite{Gal}.}
\begin{itemize}
\item integrability means the existence of $n_{\text{bos}}{+}n_{\text{fer}}$ Grassmannian-even conserved currents in involution,
\item maximal superintegrability means the existence of $2(n_{\text{bos}}{+}n_{\text{fer}}){-}1$ Grassmannian-even conserved currents.
\end{itemize}
To visualize the structure of the conserved currents we construct all Liouville charges up to level~5 for the $\cN{=}\,2$ Calogero--Moser model.
This provides explicit expressions for a complete and functionally independent set in systems with $n_{\text{bos}}\le5$. 
We advocate a general procedure and hypothesize that it generates {\it all\/} Liouville currents for an {\it arbitrary\/} number of particles in that model.
We also construct the additional set of conserved currents required for maximal superintegrability of the considered $\cN{=}\,2$ Calogero--Moser model.

\setcounter{equation}{0}
\section{Hamiltonian description}
In the Hamiltonian approach the construction of the $n$-particle rational Calogero--Moser model with $\cN$-extended supersymmetry  \cite{KLS1, KLPS} is based on the following set of components:
\begin{itemize}
\item  $n$ bosonic coordinates $x_i$ and corresponding momenta $p_i, \; i=1, \ldots,n$,
\item  $   \cN \, n^2 $ fermions $\xi^a_{ij}, \bxi_{ij\, b}$ , \quad$ {a,b =1,2,\ldots \cN/2 }$.
\end{itemize}
The non-vanishing Poisson brackets have the standard form
\be\label{PB1}
\big\{ x_i, p_j\big\}= \delta_{ij}\ , \qquad
\big\{ \xi^a_{ij}, \bxi_{km\, b}\big\} = -\im \delta^a_b\delta_{im}\delta_{jk}\ .
\ee
It is convenient to collect the bosonic coordinates in a diagonal matrix~$X$ with components
\be\label{X}
X_{ij} = \delta_{ij} x_j\ .
\ee
Basic to our construction are the fermionic bilinear objects \be\label{pi}
\Pi_{ij} = \sum_{a=1}^{\cN/2} \sum_{k=1}^n \big( \xi^a_{ik}\bxi_{kj\,a } + \bxi_{ik\,a}\xi^a_{kj}\big) \und
 \tPi_{ij} = \sum_{a=1}^{\cN/2} \sum_{k=1}^n \big( \xi^a_{ik}\bxi_{kj\,a } - \bxi_{ik\,a}\xi^a_{kj}\big)\ .
\ee
It is easily to check that they form an $s(u(n)\oplus u(n))$ algebra\footnote{Note that by definition
$\sum_i \Pi_{ii} =0$.}
\be\label{susus}
\big\{ \Pi_{ij}, \Pi_{km} \big\} = \big\{ \tPi_{ij}, \tPi_{km} \big\} = \im \big( \delta_{im} \Pi_{kj}-\delta_{kj}\Pi_{im}\big) \und
\big\{ \Pi_{ij}, \tPi_{km} \big\} = \im \big( \delta_{im} \tPi_{kj}-\delta_{kj}\tPi_{im}\big) .
\ee

The $\cN$-extended supersymmetric $A_1 \oplus A_{n-1}$ rational Calogero model
(with a harmonic confining potential) is described by supercharges~\cite{KLPS}
\be\label{Q1}
\mQ{}^a = \sum_{i=1}^n \big(p_i + \im \omega x_i \big)\xi^a_{ii}\ -\ \im \sum_{i \neq j}^n \frac{\big(g + \Pi_{jj} - \Pi_{ij}\big) \xi^a_{ji}}{x_i-x_j}\ , \quad
\bmQ{}_a =\sum_{i=1}^n \big(p_i - \im \omega x_i \big)\bxi_{ii\, a}\ +\ \im \sum_{i \neq j}^n \frac{\big(g + \Pi_{ii} - \Pi_{ji}\big) \bxi_{ij\, a}}{x_i-x_j}\ ,
\ee
and a Hamiltonian
\be\label{H1}
\mH =  \frac{1}{2}\sum_{i=1}^n p_i^2\ +\ \frac{1}{2}\sum_{i \neq j}^n \frac{\big( g+\Pi_{jj}-\Pi_{ij}\big)
	\big( g+\Pi_{ii}-\Pi_{ji}\big) }{\big(x_i-x_j\big)^2} - \frac{2}{\cal N}\,\omega\, \sum_{a=1}^{{\cal N}/2} \sum_{i,j=1}^n \xi_{ij}^a \bxi_{ji\, a} +
\frac{\omega^2}{2} \sum_{i=1}^n x_i^2\ ,
\ee
which form an $su(\frac{\cN}{2}|1)$ super-algebra together with the $R$-symmetry generators
\be\label{gen-W}
{\mathbb W}^a_b = \sum_{i,j=1}^n \xi_{ij}^a \bxi_{ji\, b} - \frac{2}{\cal N}\, \delta^a_b \sum_{c=1}^{\cN/2} \sum_{i,j=1}^n \xi_{ij}^c \bxi_{ji\, c}\ .
\ee
The R-symmetry $su\big(\frac{\cN}{2}\big)$ algebra reads
\be\label{alg-W}
\big \{{\mathbb W}^a_b, {\mathbb W}^c_d \big \} = \im \delta^a_d {\mathbb W}^c_b - \im \delta^c_b {\mathbb W}^a_d\ ,
\ee
and the remaining commutation relations of the $su\big(\frac{\cN}{2}\big|1\big)$ superalgebra are given by 
\be\label{Omega-alg}
\begin{aligned}
&\big\{ \mQ^a, \bmQ_b \big\} = - 2 \im\, \delta^a_b\, \mH + 2 \im\, \omega {\mathbb W}^a_b\ , \qquad
\big\{ \mQ^a, \mQ^b \big\}=\big\{ \bmQ_a, \bmQ_b \big\}=0 \ , \\
&\big \{ \mH, \mQ^a  \big\} = - \im\, \omega\,\frac{\cN{-}2}{\cN}\,  \mQ^a\ , \qquad
\big \{ \mH, \bmQ_a \big\} = \im\, \omega\,\frac{\cN{-}2}{\cN}\, \bmQ_a\ ,\\
&\big \{ {\mathbb W}^a_b, \mQ^c \big\} =  - \im \delta^c_b\, \mQ^a  +\im\,\frac{2}{\cN}\, \delta^a_b\, \mQ^c \ , \qquad
\big \{ {\mathbb W}^a_b, \bmQ_c \big\} =  \im \delta^a_c\, \bmQ_b - \im\,\frac{2}{\cN}\, \delta^a_b\, \bmQ_c\ .
\end{aligned}
\ee
For $\cN{=}\,4$ and $\cN{=}\,8$ this superalgebra coincides with the one given in~\cite{Sid}.
One may expect that the system \p{Q1} and \p{H1} coincides with the one considered in \cite{ref2} upon two reductions: 
the first one considered in \cite{ref1}, and the second one performed in \cite{KLPS}. 
To verify this expectation two points have to be checked: 
a) invariance of the reduction of~\cite{ref1} under the $su(\frac{\cN}{2}|1)$ superalgebra, and
b) superconformal $osp(\frac{\cN}{2}|1)$ invariance of the $\omega{=}0$ system in~\cite{ref1}.

In the limit $\omega\to0$ this turns into the $\cN$-extended super-Poincar\'{e} algebra
\be\label{N2SP}
\big\{ Q^a , \bQ_b \big\} = - 2 \im\, \delta^a_b\, H \und \big\{ Q^a, Q^b \big\}=\big\{ \bQ_a, \bQ_b \big\}=0
\ee
for the unconfined charges
\be \label{H0}
Q^a = \mQ{}^a\big|_{\omega=0}\ ,\quad
\bQ^a = \bmQ{}^a\big|_{\omega=0} \und
H = \mH\big|_{\omega=0}\ .
\ee

\setcounter{equation}0
\section{Superintegrability}
Based on the similarity of the Hamiltonian \p{H0} and the Hamiltonian of the Euler--Calogero--Moser
model~\cite{sCal} and trying to represent the supercharges as
\be\label{QL}
Q^a = \sum_{i,j=1}^n L_{ij}\xi^a_{ji} \und  \bQ_a = \sum_{i,j=1}^n L_{ij}\bxi_{ji\, a}\ ,
\ee
one may guess the Lax operator~$L$ with components \cite{KLS-N2}
\be\label{lax1}
L_{ij} = \delta_{ij}\, p_j -\im \big(1-\delta_{ij}\big) \frac{g+ \Pi_{jj}-\Pi_{ij}}{x_i-x_j}.
\ee
It is indeed easily checked that
\be
\tfrac{1}{2} \tr L^2 = H
\ee
as it should be.
For a Lax-type equation we need an associated matrix~$A$.
By a simple computation its components are found as
\be\label{M}
A_{ij} = \im \, \delta_{ij} \sum_{k \neq i}^n \frac{g+\Pi_{kk}-\Pi_{ik}}{(x_i-x_k)^2}
-\im \big( 1- \delta_{ij}\big) \frac{g+\Pi_{jj}-\Pi_{ij}}{(x_i-x_j)^2}\,.
\ee
With this, the Lax--type equations of motion related to the Hamiltonian \p{H1} read
\be\label{eomL}
\frac{\diff}{\diff t}\, L_{ij} = \big\{ L_{ij}, \mH \big\}=\big[ A, L \big]_{ij} - \omega^2 X_{ij}\ .
\ee
The equations of motion for the coordinate matrices acquire the form
\be\label{eomX}
\frac{\diff}{\diff t}\, X_{ij} =  \left\{ X_{ij}, \mH \right\}=\left[ A, X \right]_{ij}+ L_{ij} \und
\ee
\be\label{eomxi}
\frac{\diff}{\diff t}\, \xi^a_{ij} = \big\{ \xi^a_{ij}, \mH \big\}=\big[ A, \xi^a \big]_{ij} - 2 \frac{\im\,\omega}{ \cN}\, \xi^a_{ij}\ ,
\qquad
\frac{\diff}{\diff t}\,\bxi_{ij\,a}  = \big\{ \bxi_{ij\,a}, \mH \big\}=\big[ A,\bxi_a\big]_{ij}+2\frac{\im\,\omega}{\cN}\,\bxi_{ij}^a\ .
\ee
As a corollary, the composite objects $\Pi_{ij}$ and $\tPi_{ij}$ \p{pi} satisfy the following equations,
\be\label{eomPi}
\frac{\diff}{\diff t}\, \Pi_{ij} \equiv  \big\{ \Pi_{ij}, \mH \big \} = \big[A,\, \Pi \big]_{ij} \und
\frac{\diff}{\diff t}\, \tPi_{ij} \equiv  \big\{ \tPi_{ij}, \mH \big\} = \big [ A,\, \tPi  \big ]_{ij}\ .
\ee

The equations of motion \p{eomL}--\p{eomxi} are similar  to those one in~\cite{sCal,sCal1, GH}
and can be solved by the Olshanetsky--Perelomov method~\cite{OP}.
For this purpose we need an invertible time-dependent matrix $U=(U_{ij})$
as the solution of the linear differential equation
\be\label{eomU}
\frac{\diff}{\diff t}\, U_{ij} = \big(A\, U\big)_{ij} \qquad\text{with}\quad U_{ij}\big|_{t=0} = \delta_{ij}\ .
\ee
Using this matrix, one can pass to new tilded variables
\be\label{tilde}
\tilde{L}_{ij} = \big(U^{-1} L\, U\big)_{ij}\,, \quad \tilde{X}_{ij} = \big(U^{-1} X\, U\big)_{ij}\,, \quad
\tilde\xi^a_{ij} = \big(U^{-1} \xi^a\, U\big)_{ij}\,, \quad \tilde\bxi_{ij\,a} = \big(U^{-1} \bxi_a\, U\big)_{ij}\,.
\ee
In terms of these, the $A$ contribution of the equations \p{eomL}--\p{eomxi} is removed, hence
\be\label{fineomv}
\frac{\diff}{\diff t}\, \tilde{L} = - \omega^2 \tilde{X}\,, \quad 
\frac{\diff}{\diff t}\, \tilde{X} = \tilde{L} \quad\Rightarrow\quad
\frac{\diff^2}{\diff t^2}\, \tilde{X} = -\omega^2 \tilde{X}\,, \quad
\frac{\diff}{\diff t}\, \tilde{\xi}^a_{ij} =-  2 \im\, \frac{\omega}{ \cN}\, \tilde{\xi}^a_{ij}\,, \quad
\frac{\diff}{\diff t}\, \tilde{\bxi}_{ij\,a} = 2 \im \, \frac{\omega}{ \cN} \, \tilde{\bxi}_{ij\,a}\ .
\ee
The solutions to these equations are easily found as
\be\label{sol-TX}
\begin{aligned}
\tilde{L}_{ij}(t) &= \cos(\omega t) L_{ij}(0) - \omega \sin(\omega t) X_{ij}(0)\ ,\quad
\tilde{X}_{ij}(t) = \cos(\omega t) X_{ij}(0) + \omega^{-1} \sin(\omega t) L_{ij}(0) \ ,\\
\tilde{\xi}^a_{ij}(t) &= \exp(-2 \tfrac{\im\omega t}{ \cN})\, \xi^a_{ij}(0) \und
\tilde{\bxi}_{ij\,a}(t) = \exp(2 \tfrac{\im\omega t}{ \cN})\, \bxi_{ij\,a}(0)\ .
\end{aligned}
\ee
The matrix $U(t)$ then determines the time dependence of the original variables via
\be
\begin{aligned}
L_{ij}(t) &= \cos(\omega t) \bigl(U(t)L(0)U^{-1}(t)\bigr)_{ij}
- \omega \sin(\omega t) \bigl(U(t)X(0)U^{-1}(t)\bigr)_{ij} \ ,\\
X_{ij}(t) &= \cos(\omega t) \bigl(U(t)X(0)U^{-1}(t)\bigr)_{ij}
+ \omega^{-1} \sin(\omega t) \bigl(U(t)L(0)U^{-1}(t)\bigr)_{ij}  \ ,\\
\xi^a_{ij}(t) &= \exp(-2 \tfrac{\im\omega t}{ \cN})\, \bigl(U(t)\xi^a(0)U^{-1}(t)\bigr)_{ij} \und
\bxi_{ij\,a}(t) = \exp(2 \tfrac{\im\omega t}{ \cN})\, \bigl(U(t)\bxi^a(0)U^{-1}(t)\bigr)_{ij}\ .
\end{aligned}
\ee

Of course, one still has to solve \p{eomU} to obtain the entire dynamics.
However, for the time evolution of the bosonic coordinates~$x_i$ the $U$ matrix is irrelevant
because the eigenvalues of the matrix~$X$ are unchanged by the unitary transformation with~$U$.
Therefore, the motion~$x_i(t)$ is periodic, and the bosonic trajectories are closed curves in phase space.
Hence, the Hamiltonian~\p{H1} is completely degenerate with respect to its bosonic degrees of freedom,
and so these provide $2n{-}1$ functionally independent constants of motion.
This property is preserved in the unconfining limit $\omega\to0$,
implying that the $\cN$--supersymmetric Calogero--Moser model is maximally superintegrable
in its bosonic sector, just like the purely bosonic model is~\cite{sCal}.
The situation is less clear for the conserved charges formed with the fermionic degrees of freedom.
Therefore, we take a look at the explicit form of the integrals of motion.

\setcounter{equation}{0}
\section{Conserved currents}
It is interesting to know the explicit form of the integrals of motion, especially in the present
case where the number of the fermionic degrees of freedom is much larger that number of bosonic ones.
Here, we will analyze the integrability of the simplest supersymmetric Calogero--Moser model with $\cN{=}\,2$ supersymmetry.

{}From the equations of motion \p{eomL}, \p{eomxi}, \p{eomPi} it follows that
any function $\hat F$ with a polynomial dependence on the matrices $L$, $\xi^a$ or $\bxi_b$ obeys the equation
\be\label{eom}
\frac{\diff}{\diff t}\,{\hat F} = \big[ A,  {\hat F}\big]
\ee
and, therefore, the trace of such a function is conserved:
\be\label{curr1}
\frac{\diff}{\diff t}\,\tr {\hat F}(L, \xi^a, \bxi_b) = 0\ .
\ee
We note that the fermions $\xi^a$ and $\bxi_b$ have the non-standard conjugation properties~\cite{KLPS}
\be\label{con_rule}
\left( \xi^a_{ij}\right)^\dagger = \frac{g+\Pi_{jj}}{g+\Pi_{ii}}\,\bxi_{ji\, a} \qquad \Rightarrow \qquad
\left( \Pi_{ij}\right)^\dagger = \frac{g+\Pi_{jj}}{g+\Pi_{ii}}\,\Pi_{ji}\ , \qquad 
\left( L_{ij}\right)^\dagger = \frac{g+\Pi_{jj}}{g+\Pi_{ii}}\,L_{ji}\ , \qquad \textrm{etc.}
\ee 
Thus, under the trace all factors of $\frac{g+\Pi_{jj}}{g+\Pi_{ii}}$ are canceled, and we have
\be\label{tr}
\bigl(\tr {\hat F}(L, \xi^a, \bxi_b)\bigr)^\dagger = \tr {\hat F}(L, \xi^a, \bxi_b)\ .
\ee

The whole system \p{eomL}, \p{eomX}, \p{eomxi} has $2(n^2{+}n)$ dynamical variables,
i.e.~$(x_i,p_i)$ and $(\xi_{ij}, \bxi_{ij})$.\footnote{
In this section we consider the case with $\cN{=}\,2$ supersymmetry and therefore 
omit the index $`1'$ in the fermions $\xi^1_{ij}$ and $\bxi_{ij \,1}$.}
Thus the system requires $n^2+n$ functionally independent integrals of motion in the involution to be integrable.
Some of these integrals may be recovered from a spectral-parameter Lax representation~\cite{sCal}
\be\label{eom1}
\frac{\diff}{\diff t}\, \big( L+ \mu \Pi \big) = \big[ A, L+ \mu \Pi \big]\ .
\ee
The trace-powers
\be \label{int1}
\tr \big(L+ \mu \Pi \big)^k\qquad\textrm{for} \quad k=1,\ldots,n
\ee
are spectral-parameter dependent integrals of \p{eom1}. 
We expand them in powers of~$\mu$ and obtain a set~$C_n$ of conserved charges for $n$~particles.

The number~$c_n$ of integrals \p{int1} (the cardinality of~$C_n$) is given recursively by
\be
c_n = c_{n-1} +(n{+}1)\ .
\ee
Keeping in the mind that $\tr(\Pi)=0$ and, therefore, $c_1 =1$, we conclude that 
\be\label{Cn}
c_n = \tfrac12 n(n{+}1)+n-1 \ .
\ee
It is instructive to visualize the structure of these integrals for a small number of particles:
\be
\begin{aligned}
c_1{=}\ \,1 : & \quad C_1 = \left\{ \tr(L)\right\}\ , \\
c_2{=}\ \,4 : & \quad C_2 = C_1 \cup \left\{ \tr(L^2)\,,\,\tr (L\Pi)\,,\,\tr (\Pi^2) \right\}\ , \\
c_3{=}\ \,8 : & \quad C_3 = C_2 \cup \left\{ \tr(L^3)\,,\,\tr(L^2 \Pi)\,,\,\tr(L\Pi^2)\,,\;\tr(\Pi^3)\right\}\ , \\
c_4{=}13    : & \quad C_4 = C_3 \cup \left\{ \tr(L^4)\,,\,\tr (L^3 \Pi)\,,\,\tr(2 L^2 \Pi^2{+}L \Pi L\Pi)\,,\,\tr (L \Pi^3)\,,\,\tr(\Pi^4)\right\}\ , \\
c_5{=}19    : & \quad C_5 = C_4 \cup \left\{ \tr(L^5)\,,\,\tr(L^4\Pi)\,,\,\tr(L^3 \Pi^2{+}L^2 \Pi L \Pi)\,,\,\tr(L^2 \Pi^3{+}L\Pi^2 L\Pi)\,,\,\tr(L\Pi^4),\;\tr(\Pi^5)\right\} \\
\end{aligned}
\ee
and so on.
However, the number \p{Cn} of conserved currents is less that we need for the integrability, i.e.~$n^2{+}n$. 

To construct more currents one may try to use a different spectral Lax representation,
\be\label{eom2}
\frac{\diff}{\diff t} \big( L+ \mu \tPi \big) = \big[ A, L+ \mu \tPi \big]
\ee
with $\tPi$ defined in \p{pi}.
However, the integrals in the expressions
\be \label{int2}
\tr \big(L+ \mu \tPi \big)^k \qquad\textrm{for}\quad k=1,\ldots,n
\ee
obtained from expanding in powers of~$\mu$ do not commute with the currents~\p{int1}.
Thus, the remaining possibility for additional conserved currents of Liouville type resides in the expression
\be \label{int3}
\tr \big(\Pi+ \mu \tPi \big)^k \qquad\textrm{for}\quad k=1,\ldots,n\ .
\ee
We have no rigorous proof, but we checked for a small number of particles that the currents in \p{int3} perfectly commute with the currents in \p{int1}. 
Observing that the $\mu{=}0$ currents in \p{int3} already are contained in the currents \p{int1}, one evaluates the number of new currents in \p{int3} to be
\be
{\widetilde c}_n = \tfrac12 n(n+1)\ .
\ee
Thus, the total number of Liouville currents from \p{int1} and \p{int3} is $n^2{+}2n{-}1$, while
the system has $n^2{+}n$ degrees of freedom. Hence, there must exist $n{-}1$ constraints among the currents~\p{int3}.
One may check that these constraints read
\be\label{finalC}
2\,\chi_k \ \equiv\ \tr\big(\Pi + \tPi \big)^k + \tr\big(\Pi - \tPi \big)^k\ =\ 0
%\chi_k(\mu)\ \equiv\ \tr\Big[\big(\mathbbm{1} + \mu \big(\Pi + \tPi \big)\big)^k \Big] 
%+ \tr\Big[ \big(\mathbbm{1} + \mu \big(\Pi - \tPi \big) \big)^k\Big] -2n\ =\ 0
\qquad\textrm{for}\quad k=1,\ldots, n\ .
\ee
Until level 5 the currents look as follows,
\be\label{relation2}
\begin{aligned}
\chi_1 &\ =\ \tr (\Pi) \ , \\
\chi_2 &\ =\ \tr ( \Pi^2 ) + \tr ( {\tPi}^2 ) \ , \\
\chi_3 &\ =\ \tr ( \Pi^3 ) + 3 \tr ( \Pi {\tPi}^2 ) \ , \\
\chi_4 &\ =\ \tr ( \Pi^4 ) + 4 \tr ( \Pi^2 {\tPi}^2 ) + 2 \tr ( \Pi \tPi \Pi \tPi ) + \tr ( {\tPi}^4 ) \ , \\
\chi_5 &\ =\ \tr ( \Pi^5 ) + 5 \tr ( \Pi^3 {\tPi}^2 ) + 5 \tr ( {\Pi}^2 \tPi \Pi \tPi ) + 5 \tr ( \Pi {\tPi}^4 ) \ . 
\end{aligned}
\ee
A set of ${\widetilde c}_k$ independent additional Liouville integrals up to this level is
\be
\begin{aligned}
{\widetilde c}_1{=}\ \,1 : & \quad {\widetilde C}_1 = \big\{ \tr(\tPi)\big\}\ , \\
{\widetilde c}_2{=}\ \,2 : & \quad {\widetilde C}_2 = {\widetilde C}_1 \cup \big\{ \tr (\Pi\tPi) \big\} \ , \\
{\widetilde c}_3{=}\ \,4 : & \quad {\widetilde C}_3 = {\widetilde C}_2 \cup \big\{ \tr(\tPi^3)\,,\,\tr(\tPi\Pi^2)\big\} \ , \\
{\widetilde c}_4{=}\ \,7 : & \quad {\widetilde C}_4 = {\widetilde C}_3 \cup \big\{ \tr (\tPi^3 \Pi)\,,\,\tr(2 \tPi^2 \Pi^2{+}\tPi \Pi \tPi\Pi)\,,\,\tr (\tPi \Pi^3)\big\}\ ,\\
{\widetilde c}_5{=}11    : & \quad {\widetilde C}_5 = {\widetilde C}_4 \cup \big\{ \tr(\tPi^5)\,,\,\tr(\tPi^3 \Pi^2{+}\tPi^2 \Pi \tPi \Pi)\,,\,\tr(\tPi^2 \Pi^3{+}\tPi\Pi^2 \tPi\Pi)\,,\,\tr(\tPi\Pi^4)\big\}\ .
\end{aligned}
\ee

The construction of the non-involutive conserved currents needed for superintegrability
of the $\cN{=}\,2$ Calogero--Moser model proceeds in full analogy with the one discussed
in~\cite{DLM1,DLM2,CLP} for the standard $\cN{=}\,2$ supersymmetric Calogero--Moser model \cite{FM}.
Our Liouville charges all arise from some matrix $\cB=(\cB_{ij})$ with an equation of motion
\be\label{eomB}
\frac{\diff}{\diff t}\, \cB_{ij}\ =\ \big[ A, \cB \big]_{ij} \qquad\Rightarrow\qquad
\frac{\diff}{\diff t}\,\tr (\cB^k) = 0\ .
\ee
One way to derive further conserved charges is to ``dress'' $\cB$ with the matrix~$X$, but
\be
\frac{\diff}{\diff t}\, \tr (X\cB)= \tr (L\cB)
\ee
due to~\p{eomX}. However, the right-hand side cancels in the antisymmetric two-trace expression
\be\label{wB}
\widetilde{\cB} = \tr\big(X \cB\big) \tr \big( L \big)  - \tr\big( L \cB\big) \tr\big( X\big) 
\qquad\Rightarrow \qquad \frac{\diff}{\diff t}\, {\widetilde\cB} = 0\ .
\ee
Using the matrices $(L+\mu \Pi)^k$ from~\p{int1} in place of $\cB$
one may construct $\tfrac12 n(n{+}1){-}1$ new currents of the type~\p{wB} while, using
the matrices  $(\Pi+\mu \widetilde{\Pi})^k$ from~\p{int3}, we may produce  $\tfrac12 n(n{+}1)$ new currents.
Thus, for the system with $n(n{{+}}1)$ degrees of freedom 
we obtain in total $2n(n{{+}}1){-}1$ conserved currents.
This is the correct count indeed rendering the $\cN{=}\,2$ supersymmetric Calogero--Moser model 
maximally superintegrable.

\section{Conclusion}
We have analyzed the integrability of the $\cN$-extended supersymmetric Calogero--Moser model. 	
We explicitly constructed the Lax pair for this system and proved maximal superintegrability,
at least for the bosonic sector. A procedure has been proposed for constructing sufficiently many
functionally independent conserved currents, including but extending the Liouville charges.
We have demonstrated this by explicit expressions up to level five.

One may try to repeat this analysis for the $\cN$-extended supersymmetric Euler--Calogero--Moser \cite{KLS2}
and Calogero--Moser--Sutherland \cite{KL2}  models. 
The generalization of all models from $A_{n-1}$ to other Coxeter groups may also be investigated 
in view of possible integrability though this is less straightforward.

\vspace{0.5cm}

\noindent{\bf Acknowledgements}\\
The work was supported by Russian Foundation for Basic Research, grant no.~20-52-12003.\\
\noindent
We thank to Armen Nersessian for stimulating discussions.

%\newpage


\begin{thebibliography}{99}
%\addtolength{\itemsep}{-3pt}
\bibitem{Calogero} F.~Calogero, \\
{\it Ground state of a one-dimensional N-body system},\\
J.~Math.~Phys. {\bf 10} (1969) 2179;\\
{\it Solution of a three-body problem in one dimension},\\
J.~Math.~Phys. {\bf 10} (1969) 2191; \\
{\it Solution of the one-dimensional N-body problems with quadratic and/or inversely quadratic pair potentials},\\
J.~Math.~Phys. {\bf 12} (1971) 419. 
\bibitem{OP} M.A.~Olshanetsky, A.M.~Perelomov, \\
{\it Explicit solution of the Calogero model in the classical case and geodesic flows on symmetric spaces of zero curvature}, \\
Lett.~Nuovo Cimento {\bf 16} (1976) 333.
\bibitem{FM} D.Z.~Freedman, P.F.~Mende, \\
{\it An exactly solvable N-particle system in supersymmetric quantum mechanics},\\
Nucl.~Phys.~B {\bf 344} (1990) 317.
\bibitem{GT} G.W.~Gibbons, P.K.~Townsend,\\
{\it Black holes and Calogero models}, \\
Phys.~Lett.~B {\bf 454} (1999) 187, {\tt arXiv:hep-th/9812034}.
\bibitem{Wyll} N.~Wyllard, \\
{\it (Super)conformal many-body quantum mechanics with extended supersymmetry}, \\
J.~Math.~Phys. {\bf 41} (2000) 2826, {\tt arXiv:hep-th/9910160}.
\bibitem{BGL} S.~Bellucci, A.~Galajinsky, E.~Latini,\\
{\it New insight into the Witten--Dijkgraaf--Verlinde--Verlinde equation},\\
Phys.~Rev.~D {\bf 71} (2005) 044023, {\tt arXiv:hep-th/0411232}.
\bibitem{scm8} S.~Fedoruk, E.~Ivanov, O.~Lechtenfeld,\\
{\it Supersymmetric Calogero models by gauging},\\
Phys.~Rev.~D {\bf 79} (2009) 105015, {\tt arXiv:0812.4276[hep-th]}. 
\bibitem{nCal3} S.~Fedoruk, E.~Ivanov, O.~Lechtenfeld,\\
{\it OSp$(4|2)$ superconformal mechanics},\\
JHEP {\bf 08} (2009) 081,  {\tt arXiv:0905.4951[hep-th]}. 
\bibitem{nCal2}S.~Fedoruk, E.~Ivanov, O.~Lechtenfeld,\\
{\it New $D(2,1;\alpha)$ mechanics with spin variables},\\
JHEP {\bf 04} (2010) 129 , {\tt arXiv:0912.3508[hep-th]}.
\bibitem{nCal1}S.~Fedoruk, E.~Ivanov, O.~Lechtenfeld,\\
{\it New super Calogero models and OSp$(4|2)$ superconformal mechanics},\\
Phys.~Atom.~Nucl. {\bf 74} (2011) 870, {\tt arXiv:1001.2536[hep-th]}. 
\bibitem{KLS1}S.~Krivonos, O.~Lechtenfeld, A.~Sutulin,\\
{\it $\cN$-extended supersymmetric Calogero model},\\
Phys.~Lett.~B {\bf 784} (2018) 137, {\tt arXiv:1804.10825[hep-th]}.
\bibitem{KLPS}S.~Krivonos, O.~Lechtenfeld, A.~Provorov, A.~Sutulin,\\
{\it Extended supersymmetric Calogero model}, \\
Phys.~Lett.~B {\bf 791} (2019) 385, {\tt arXiv:1812.10168[hep-th]}.
\bibitem{sCal} S.~Wojciechowski,\\
{\it An integrable marriage of the Euler equations with the Calogero--Moser system},\\
Phys.~Lett.~A {\bf 111} (1985) 101.
\bibitem{DLM1} P.~Desrosiers, L.~Lapointe, P.~Mathieu, \\
{\it Supersymmetric Calogero--Moser--Sutherland models and Jack superpolynomials},\\
Nucl.~Phys.~B {\bf 606} (2001) 547, {\tt arXive:hep-th/0103178}.
\bibitem{DLM2} P.~Desrosiers, L.~Lapointe, P.~Mathieu, \\
{\it Supersymmetric Calogero--Moser--Sutherland models: superintegrability structure and eigenfunctions},\\
in: proceedings of the Workshop on Superintegrability in Classical and Quantum Systems,\\
16-22 Sept 2002, Montreal, Quebec, Canada,  {\tt arXive:hep-th/0210190}.
\bibitem{Gal} A.~Galajinsky,\\
{\it Remarks on N=1 supersymmetric extension of the Euler top},\\
Nucl.~Phys.~B {\bf 975} (2022) 115668, {\tt arXiv:2111.06083}.
\bibitem{Sid} S.~Sidorov, \\
{\it Hidden supersymmetries of deformed supersymmetric mechanics},\\
in: proceedings of ISQS-26,
J.~Phys.~Conf.~Ser.~{\bf 1416} (2019) 012032, {\tt arXiv:1910.00198[hep-th]}.
\bibitem{ref2}  S.~Fedoruk, E.~Ivanov, S.~Sidorov, \\
{\it Quantum SU$(2|1)$ supersymmetric Calogero--Moser spinning systems}, \\
JHEP {\bf 04} (2018) 043, {\tt arXiv:1801.00206[hep-th]}.
\bibitem{ref1} S.~Fedoruk, \\
{\it $\cN{=}\,4$ supersymmetric U(2)-spin hyperbolic Calogero--Sutherland model}, \\
Nucl.~Phys.~B {\bf 961} (2020) 115234, {\tt arXiv:2007.11424[hep-th]}.
\bibitem{KLS-N2} S.~Krivonos, O.~Lechtenfeld, A.~Sutulin,\\
{\it New $\cN{=}\,2$ superspace Calogero model},\\
JHEP {\bf 05} (2020) 132, {\tt  arXiv:1912.05989[hep-th]}.
\bibitem{sCal1} S.~Wojciechowski,\\
{\it Superintegrability of the Calogero--Moser system},\\
Phys.~Lett. {\bf A 95} (1983) 279.
\bibitem{GH} J.~Gibbons, T.~Hermsen, \\
{A generalisation of the Calogero--Moser system}, \\
Physica D {\bf 11} (1984) 337.
\bibitem{CLP}
C.~Correa, O.~Lechtenfeld, M.~Plyushchay,\\
{\it Nonlinear supersymmetry in the quantum Calogero model},\\
JHEP {\bf 04} (2014) 151, {\tt arXiv:1312.5749[hep-th]}.
\bibitem{KLS2} S.~Krivonos, O.~Lechtenfeld, A.~Sutulin,\\
{\it Supersymmetric many-body Euler--Calogero--Moser model}, \\
Phys.~Lett.~B {\bf 790} (2019) 191, {\tt arXiv:1812.03530[hep-th]}.
\bibitem{KL2}  S.~Krivonos, O.~Lechtenfeld, \\
{\it $\cN{=}\,4$  supersymmetric Calogero--Sutherland models},\\
Phys.~Rev.~D {\bf 101} (2020) 086010, {\tt arXiv:2002.03929[hep-th]}.
\end{thebibliography}
\end{document}